\documentclass[twocolumn,prb,showpacs,preprintnumbers,amsmath,amssymb,superscriptaddress]{revtex4-1}

\usepackage[nottoc,numbib]{tocbibind}

\usepackage[colorlinks=true, citecolor=black, linkcolor=black, urlcolor=black]{hyperref}

\usepackage[all]{hypcap}
\usepackage{amsmath}
\usepackage{enumerate,bm}
\usepackage{graphicx}
\usepackage{grffile} 
\usepackage{color}
\usepackage{esint}
\usepackage{textpos}
\usepackage[usenames,dvipsnames]{xcolor}
\usepackage{subfigure}
\usepackage{flushend}
\usepackage{tabularx}
\usepackage{bbold}

\usepackage{amssymb}
\usepackage{float}
\usepackage{subfigure}
\usepackage{ulem} 

\begin{document}

\author{M.~A.~Sentef}
\email[]{michael.sentef@mpsd.mpg.de}
\affiliation{Max Planck Institute for the Structure and Dynamics of Matter,
Center for Free Electron Laser Science, 22761 Hamburg, Germany}


\title{
Light-enhanced electron-phonon coupling from nonlinear electron-phonon coupling
}
\date{\today}
\begin{abstract}
We investigate an exact nonequilibrium solution of a two-site electron-phonon model, where an infrared-active phonon that is nonlinearly coupled to the electrons is driven by a laser field. 
The time-resolved electronic spectrum shows coherence-incoherence spectral weight transfer, a clear signature of light-enhanced electron-phonon coupling. The present study is motivated by recent evidence for enhanced electron-phonon coupling in pump-probe TeraHertz and angle-resolved photoemission spectroscopy in bilayer graphene when driven near resonance with an infrared-active phonon mode [E.~Pomarico et al., Phys.~Rev.~B 95, 024304 (2017)], and by a theoretical study suggesting that transient electronic attraction arises from nonlinear electron-phonon coupling [D.~M.~Kennes et al., Nature Physics (2017), 10.1038/nphys4024]. We show that a linear scaling of light-enhanced electron-phonon coupling with the pump field intensity emerges, in accordance with a time-nonlocal self-energy based on a mean-field decoupling using quasi-classical phonon coherent states. Finally we demonstrate that this leads to enhanced double occupancies in accordance with an effective electron-electron attraction. Our results suggest that materials with strong phonon nonlinearities provide an ideal playground to achieve light-enhanced electron-phonon coupling and possibly light-induced superconductivity.
\end{abstract}
\maketitle

\section{Introduction}
Resonant laser excitation of phonons in crystal lattices allows for mode-specific stimulations of coherent lattice motion with large amplitudes. In pump-probe experiments, transient rectification effects on electronic properties have been suggested based on the notion of nonlinear phonon-phonon couplings, that is, lattice anharmonicities, an idea coined ``nonlinear phononics'' \cite{forst_nonlinear_2011, subedi_theory_2014}. Lattice excitation has been shown to lead to ultrafast metal-insulator transitions \cite{rini_control_2007, caviglia_ultrafast_2012}, control of magnetism \cite{forst_driving_2011, forst_spatially_2015, nova_effective_2017}, or enhanced superconducting-like optical signatures \cite{kaiser_optically_2014, hu_optically_2014, mitrano_possible_2016}. Recently signatures of enhanced transient electron-phonon coupling were reported in driven bilayer graphene \cite{pomarico_enhanced_2017}. The dynamical modulation of effective electronic parameters \cite{dunlap_dynamic_1986} has been discussed in various contexts ranging from bond softening in molecules \cite{bucksbaum_softening_1990} to quantum dots \cite{creffield_ac-driven_2002}, atomic gases \cite{eckardt_superfluid-insulator_2005, hemmerich_effective_2010, coulthard_enhancement_2016}, spin models, \cite{mentink_ultrafast_2015}, electronically driven electron-phonon systems \cite{sayyad_coexistence_2015, dutreix_dynamical_2017}, nonequilibrium superconductivity \cite{sentef_theory_2016, knap_dynamical_2016, komnik_bcs_2016, sentef_theory_2017, babadi_theory_2017, murakami_nonequilibrium_2017, mazza_non-equilibrium_2017}, or Floquet systems with modified topological properties \cite{oka_photovoltaic_2009, lindner_floquet_2011, jotzu_experimental_2014, sentef_theory_2015, hubener_creating_2017}. 

Here it is shown that nonlinear electron-phonon couplings of infrared-active phonon modes provide a route to light-enhanced electron-phonon coupling. Nonlinear electron-phonon couplings of Raman-active modes have been prominently featured in MgB$_2$ \cite{yildirim_giant_2001} and investigated theoretically with respect to equilibrium properties in Refs.~\onlinecite{adolphs_going_2013, li_effects_2015, li_quasiparticle_2015}. Nonlinear couplings of infrared-active modes have recently been suggested to possibly lead to transient electron-electron attraction by electronic squeezing of phonons, an effect that can be understood via a time-local unitary transformation of the instantaneous Hamiltonian \cite{kennes_transient_2017}. It was shown by Kennes et al.~that a quadratic electron-phonon coupling implies an electron-density-dependent shift of the bosonic oscillator stiffness that in turn induces electron-electron attraction. Here we propose that time-nonlocal correlations and dynamical spectral weight transfer in time-resolved spectroscopies are key to understanding how nonlinear electron-phonon couplings directly enhance electron-phonon couplings for driven phonons, and thereby ultimately also lead to transient electronic attraction in pump-probe experiments near a phonon resonance. To this end, we compute explicitly the exact time propagation of a driven two-site model and analyze the light-induced production of incoherent spectral weight quantitatively. From the scaling with the driving field intensity, we argue that an effective nonequilibrium self-energy to second order in the coupling and the field describes the essential effect. In addition, we show that the local double occupancy is enhanced in the driven system, providing further evidence that a light-enhanced electron-phonon coupling may indeed lead to light-induced attraction and ultimately superconducting pairing. The role of effective disorder from bosonic coherence leading to light-induced localization, and light-induced electronic excitations, are also discussed.

The paper is organized as follows: The model and method are introduced in Sec.~\ref{sec:model}. Results are presented in Sec.~\ref{sec:results}: Light-induced spectral weight transfer (Sec.~\ref{sec:transfer}), the scaling of light-enhanced electron-phonon coupling with the external field (Sec.~\ref{sec:scaling}), the light-induced effective attraction measured by the double occupancy (Sec.~\ref{sec:attraction}), and the role of electronic excitations measured by the occupied part of the electronic spectrum (Sec.~\ref{sec:excitations}). Conclusions are drawn in Sec.~\ref{sec:conclusion}.

\section{Model and Method}
\label{sec:model}
We consider the driven half-filled two-site model with two electrons (one spin-up, one spin-down), studied previously without the driving field and shown to lead to induced electronic attraction when starting from a bosonic coherent state in Ref.~\onlinecite{kennes_transient_2017},
\begin{align}
\hat{H}(t) &= -J \sum_{\sigma} (c^{\dagger}_{1,\sigma} c^{}_{2,\sigma} + c^{\dagger}_{2,\sigma} c^{}_{1,\sigma}) 
\nonumber \\
& + 
g_2 \sum_{\sigma, l=1,2} \hat{n}_{l,\sigma} (b^{}_{l}+b^{\dagger}_l)^2 \nonumberÊ\\
&+ \Omega \sum_{l=1,2} b^{\dagger}_l b^{}_l + F(t) \sum_{l=1,2} (b^{}_{l}+b^{\dagger}_l),
\end{align} 
where $J$ is the electronic hopping matrix element between sites $l=1,2$, $c^{(\dagger)}_{l,\sigma}$ annihilates (creates) an electron of spin $\sigma=\uparrow, \downarrow$ on site $l$ with electron number operator $\hat{n}_{l,\sigma} \equiv c^{\dagger}_{l,\sigma} c^{}_{l,\sigma}$, 
$g_2$ is the nonlinear electron-phonon coupling with bosonic phonon annihilation (creation) operators $b^{(\dagger)}_l$ on site $l$, $\Omega$ is the phonon frequency, and $F(t)$ is a driving field coupling to the phononic position (dipole) operator $\hat{x}_l \equiv b^{}_{l}+b^{\dagger}_l$. In a generic lattice with inversion symmetry, the phonon modes to which $F(t)$ couples are ungerade and thus infrared active, which does {\it not} allow for a linear term and makes the $g_2$ term the lowest order allowed interaction term for infrared phonons. By multiplying out the $(b^{}_{l}+b^{\dagger}_l)^2$ term, one finds that the nonlinear interaction renormalizes the phonon frequency locally on site $l$ to $\Omega_{\text{eff}} \equiv \Omega + 2 g_2 \langle \hat{n}_l\rangle$, where $\langle \hat{n}_l \rangle \equiv \langle \hat{n}_{l,\uparrow} + \hat{n}_{l,\downarrow} \rangle = 1$ is the average local electronic occupation with electron number operator $\hat{n}_{l,\sigma} \equiv c^{\dagger}_{l,\sigma} c^{}_{l,\sigma}$. The system is driven out of equilibrium by a time-dependent periodic field 
\begin{align}
F(t) &= F \sin(\omega t),
\end{align}
with laser frequency $\omega$. 

The time-evolved wave function of the system is computed by starting in the ground state $|\psi_0\rangle$ at time $t=0$ and propagating forward in time,
\begin{align}
|\psi(t)\rangle &= \mathcal{T} e^{-i \int_{0}^{t} H(t') dt'} |\psi_0\rangle.
\end{align}
In practice we use the commutator-free fourth order scheme introduced in Ref.~\onlinecite{alvermann_high-order_2011} to compute the time-ordered ($\mathcal{T}$) exponentials on a discretized time grid for stepping from $t$ to $t+\delta_t$, 
\begin{align}
\mathcal{T} e^{-i \int_{t}^{t+\delta_t} H(t') dt'} &\approx e^{-i (c_1 H_1 + c_2 H_2) \delta_t} e^{-i (c_2 H_1 + c_1 H_2) \delta_t}, \\ 
c_{1/2} = \frac{3 \mp 2 \sqrt{3}}{12}, \; & H_{1/2} = H(t+(1/2 \mp \sqrt{3}/6)\delta_t)\nonumber,
\end{align}
which is accurate up to order $\delta_t^4$ in the time step $\delta_t$ while still avoiding the computation of commutator correction terms for a time-varying Hamiltonian. This allows us to use a relatively coarse time step of $\delta_t = 0.5$, notably much smaller than the laser field oscillation period ($t_{\omega} = 2 \pi/\omega \approx 11.4$ for $\omega = 0.55$), without time discretization issues. Convergence in the time step size was checked. The phononic Hilbert space is truncated using a cutoff of up to 23 phonons per site for the strongest fields, and convergence in the phonon cutoff checked. Signatures of electron-phonon coupling in the electronic single-particle spectrum are extracted by computing the time-resolved electronic spectrum \cite{freericks_theoretical_2009} for site $1$ and spin $\uparrow$ with spectral intensity
\begin{align}
I(\omega, t_0) &= \text{Re} \int dt_1 \; dt_2 \; e^{i \omega (t_1 - t_2)} s_{t_1,t_2,\tau}(t_0)  \nonumber \\ &\times \Big[  \langle \psi(t_2) | c^{\dagger}_{1,\uparrow} \mathcal{T} e^{-i \int_{t_1}^{t_2} H(t) dt} c^{}_{1,\uparrow} | \psi(t_1) \rangle + \nonumber \\  & \;\;\; + \langle \psi(t_1) | c^{}_{1,\uparrow} \mathcal{T} e^{-i \int_{t_2}^{t_1} H(t) dt} c^{\dagger}_{1,\uparrow} | \psi(t_2) \rangle \Big],
\label{eq:spectrum}
\end{align}
using the retarded Green's function. The second and third lines in Eq.~(\ref{eq:spectrum}) are the lesser and greater Green's functions, which contain information about the occupied and unoccupied spectral intensities, respectively. We employ a Gaussian probe pulse shape function 
\begin{align}
s_{t_1,t_2,\sigma}(t_0) \equiv \frac{1}{2 \pi \sigma} e^{-\frac{(t_1-t_0)^2}{2 \sigma^2}} e^{-\frac{(t_2-t_0)^2}{2 \sigma^2}}
\end{align}
centered around probe time $t_0$ with probe duration $\sigma$. The duration of the probe pulse plays the role of the time scale of an effective degree of freedom that ``sees'' signatures of effective couplings out of equilibrium.

We set $J=0.15$, $\Omega=0.5$, and $g_2=-0.05$ and propagate the wavefunction from $t=0$ to $t=50$. For the given parameters the renormalized phonon frequency is $\Omega_{\text{eff}} = 0.40$. We note that this choice of negative $g_2$ is not mandatory, and the light-induced spectral weight transfer discussed in the following can be observed for positive $g_2$ as well. However, the present case is particularly interesting, as the phonon is softened by the coupling to the electrons, leading to a static instability of the crystal lattice for $\Omega + 4 g_2 < 0$ at electronic half-filling, i.e.~by triggering alternating empty and double occupations that locally lead to negative phonon frequencies on the doubly-occupied sites. In principle strong driving could therefore lead to a dynamical lattice instability in this case, which is a subject for further investigations. For the time-resolved spectroscopy, the probe duration is taken to be $\sigma=8$ with center time $t_0 = 25$. 

\section{Results}
\label{sec:results}

\subsection{Spectral weight transfer}
\label{sec:transfer}

\begin{figure}[ht!pb]
\includegraphics[clip=true, trim=0 0 0 0,width=0.8\columnwidth]{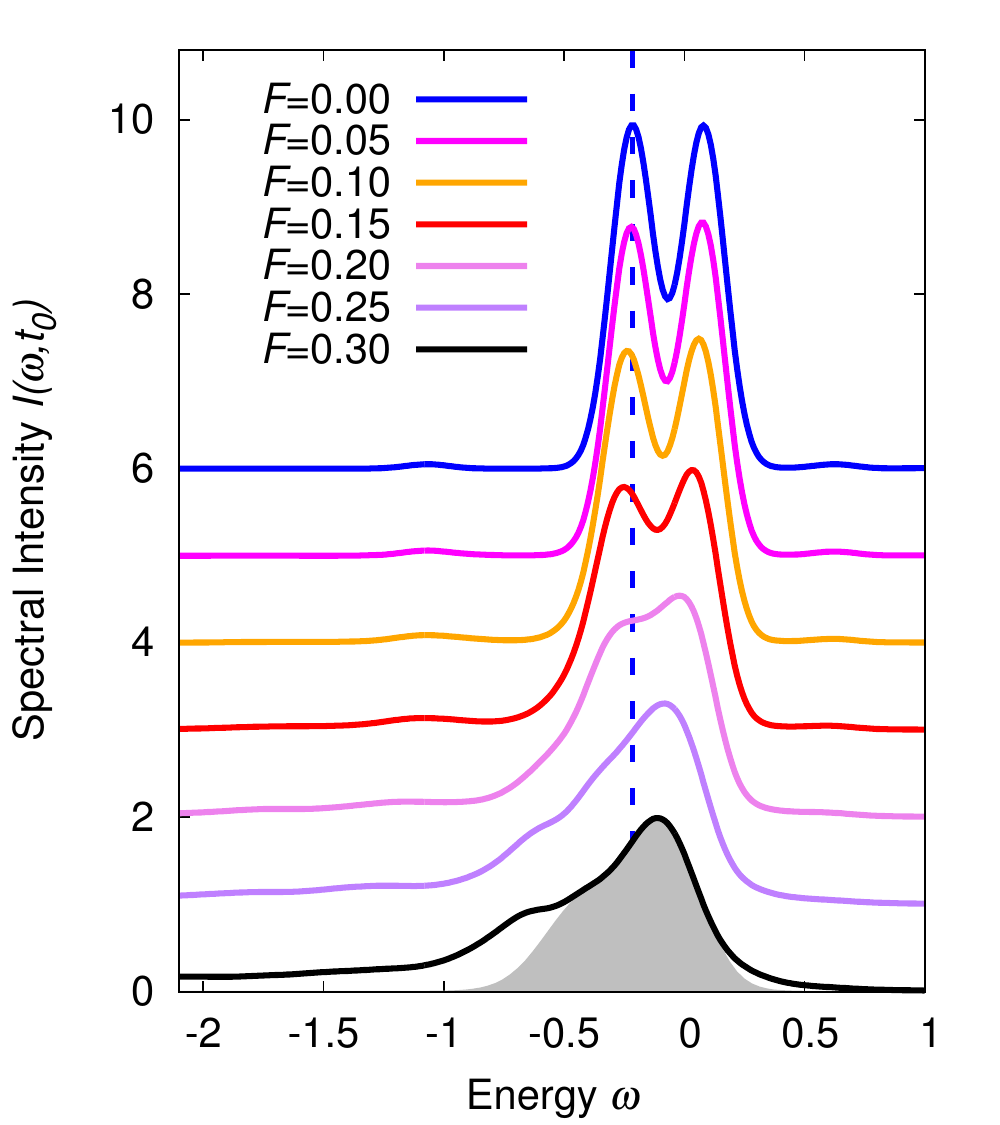}
\caption{
{\bf Light-induced spectral weight transfer.} 
Single-particle spectrum for $g_2=-0.05$ and different field strengths $F$, as indicated. Spectra are shifted vertically for clarity. The dashed vertical line indicates the peak position of the bonding state in the undriven case. For the lowest curve, the coherent part of the spectrum that emerges from the main peaks at weak driving field is indicated by the grey-shaded area.}
\label{fig1}
\end{figure}

Fig.~\ref{fig1} shows the spectral intensity during laser irradiation with slightly off-resonant field frequency $\omega=0.55$ and field strengths $F = 0.00 \dots 0.30$ as indicated. In the undriven case (top), there are dominant spectral lines corresponding to the bonding and antibonding states, with energy position of the bonding state indicated by the vertical dashed line. One can also see faint two-phonon sidepeaks roughly $2 \Omega_{\text{eff}} = 0.80$ below and above the main peaks, respectively. As the field is turned on, the main peaks broaden and lose spectral weight. At the same time they also shift down in energy. This line shift stems mainly from the local electronic energy contribution $g_2 \hat{n}_l \langle 2 b^{\dagger}_l b^{}_l + 1 \rangle < 0$ (for $g_2 < 0$), which increases in magnitude approximately linearly with $F$, as more energy is pumped into the phonons when $F$ increases. For the strongest drivings, one clearly sees the emergence of incoherent spectral weight and strongly reduced coherent peaks, indicating dynamical polaron formation via spectral weight transfer. By varying the driving frequency, we have checked that the additional peaks in the incoherent part of the spectrum are not Floquet sidepeaks \cite{sentef_theory_2015} but really incoherent spectral weight related to electron-phonon coupling. We also note that spectral redistribution in pump-probe experiments was investigated in Refs.~\onlinecite{kemper_effect_2014} and \onlinecite{rameau_energy_2016} for electronically driven systems. In stark contrast to the present work, it was found that electronically driven systems usually look ``less correlated'' rather than ``more correlated'' compared to thermal equilibrium.

\subsection{Field scaling}
\label{sec:scaling}

\begin{figure}[ht!pb]
\includegraphics[clip=true, trim=0 0 0 0,width=\columnwidth]{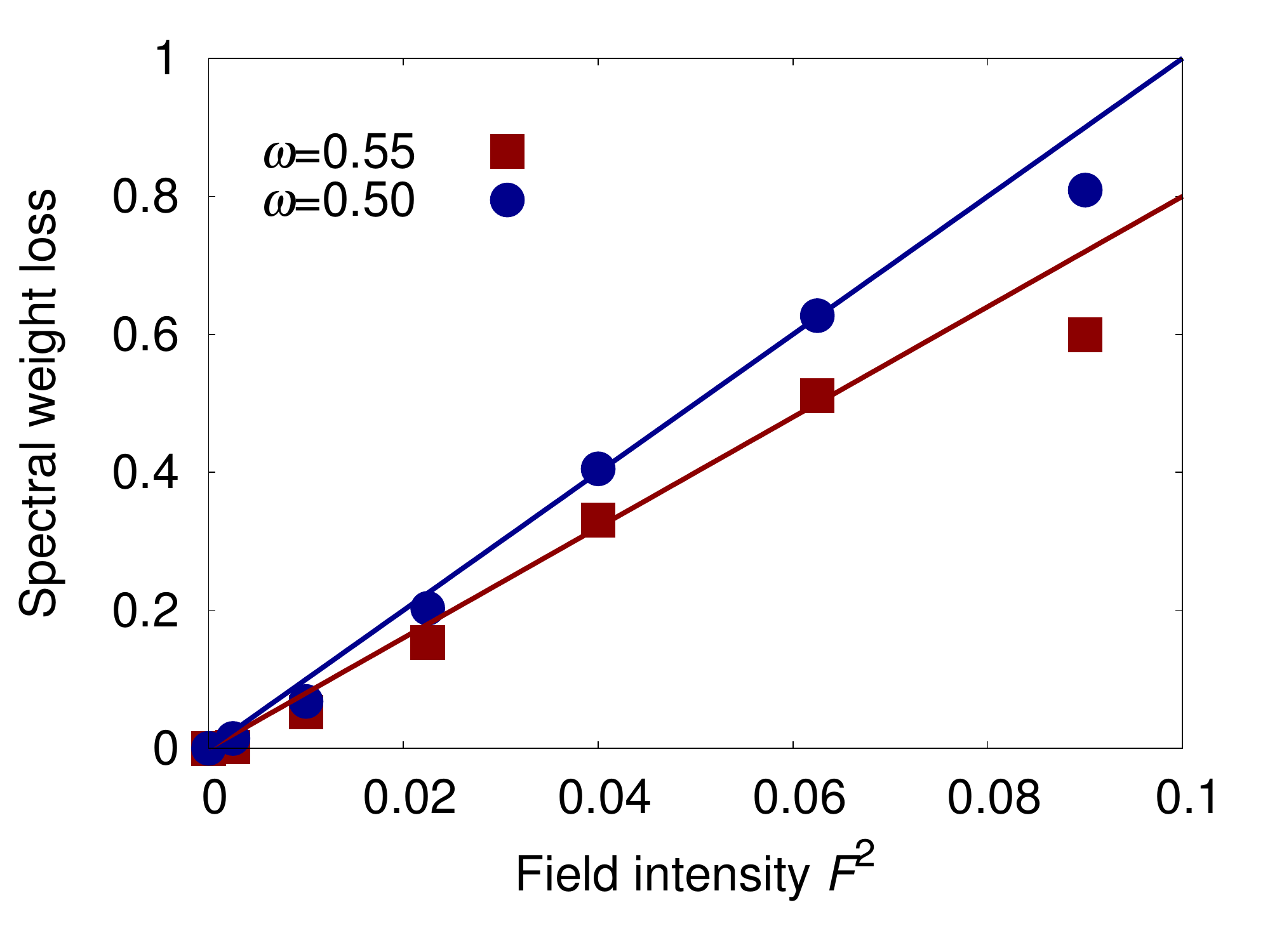}
\caption{
{\bf Field intensity scaling.} 
Spectral weight loss in the coherent peaks, extracted by fitting a sum of Gaussians to the corresponding peaks, as a function of field intensity $F^2$ at fixed $g_2=-0.05$ and two different driving frequencies $\omega=0.55$ and $\omega=0.50$, respectively. The straight lines are guides to the eye.}
\label{fig2}
\end{figure}

Having demonstrated that a driven nonlinearly coupled phonon leads to coherence-incoherence phenomena in the time-resolved electronic spectra, we now investigate quantitative aspects of the laser-induced spectral redistribution. To this end, we fit a pair of Gaussians to the coherent part of the spectrum, as shown in one example in Fig.~\ref{fig1} for the lowest curve. We subtract the fitted spectral weight from the one at $F=0$ and obtain the spectral weight loss shown in Fig.~\ref{fig2} as a function of the pump field intensity $F^2$ for two different driving frequencies. This spectral weight loss is proportional to the coherent quasiparticle weight ($Z$) that is renormalized from its value $Z_0$ at $F=0$. Using that the loss $Z_0-Z$ is proportional to $\lambda - \lambda_0$, the enhancement of effective dimensionless electron-phonon coupling $\lambda$, a proportionality that holds at weak coupling. Apparently Fig.~\ref{fig2} suggests
\begin{align}
\lambda - \lambda_0 &\propto F^2,
\end{align}
where $\lambda_0$ is the dimensionless electron-phonon coupling at zero field. Only at the strongest fields considered, we observe a saturation effect deviating from linear behavior, which is expected since the maximal spectral weight loss is bounded (for two electrons, $2Z \in [0,2]$) and the linear behavior of the spectral weight loss with effective $\lambda$ only holds at small $\lambda$. Keeping in mind the uncertainty that comes with the fitting of quasiparticle spectral weight, the linear scaling at not too strong fields leads us to predict a linear scaling of light-enhanced electron-phonon coupling with the driving field intensity or, equivalently, the pump fluence in a pump-probe experiment. One can also see in Fig.~\ref{fig2} that the effect is stronger as the driving frequency $\omega$ moves closer to the resonance frequency $\Omega_{\text{eff}}=0.40$. 

We now seek a minimal explanation for the observed scaling behavior. To this end, we first notice that a driven mode is expected to approach coherent state with well-defined phonon coordinate exhibiting quasi-classical forced oscillations $\langle \hat{x}_l(t) \rangle \propto F \sin(\omega t)$, described by boson coherent states. A mean-field decoupling yields $\Omega_{\text{eff}} = \Omega + 2 g_2 \langle \hat{n}_l \rangle$ and an interaction term $g_2 \hat{n}_l (b^{}_{l} \langle b^{}_l (t) \rangle+b^{\dagger}_l \langle b^{\dagger}_l (t) \rangle$, with oscillating mean fields $\langle b^{}_l (t) \rangle$ and $\langle b^{\dagger}_l (t) \rangle$ such that $\langle b^{}_l (t) + b^{\dagger}_l (t)  \rangle \propto F \sin(\omega t)$. In the mean-field picture, the interaction looks like a linear interaction with a \textit{time-dependent interaction vertex} that scales linearly in $g_2$, and via the coherent-phonon mean fields also linearly in $F$. 

In many-body perturbation theory, the lowest-order time-nonlocal self-energy contribution is the first Born approximation, or Migdal diagram, 
\begin{align}
\Sigma(t,t') &= i g(t) g^*(t') G(t,t') D(t,t'),
\end{align}
where we have dropped site and spin indices and introduced the local electronic Green's function $G(t,t') \equiv -i \langle \mathcal{T}_{\mathcal{C}} c^{}(t) c^{\dagger}(t') \rangle$ and phonon Green's function $D(t,t') \equiv -i \langle \mathcal{T}_{\mathcal{C}} \hat{x}(t) \hat{x}(t') \rangle$ on the three-branch Kadanoff-Baym-Keldysh contour $\mathcal{C}$ with contour-time ordering $\mathcal{T}_{\mathcal{C}}$. From this Migdal diagram one can see that the above $F^2$ scaling is indeed explained via the $F^2$ scaling of the pair of time-dependent vertices $g(t) g^*(t')$. We notice that this interpretation of enhanced electron-phonon coupling via a time-nonlocal self-energy is quite natural, but somewhat different from the time-local interpretation in Ref.~\onlinecite{kennes_transient_2017} using a unitary squeezing transformation. Here we have shown that this self-energy provides a consistent picture for a quantitative understanding of the light-induced spectral weight transfer.

\subsection{Effective attraction}
\label{sec:attraction}

\begin{figure}[ht!pb]
\includegraphics[clip=true, trim=0 0 0 0,width=0.9\columnwidth]{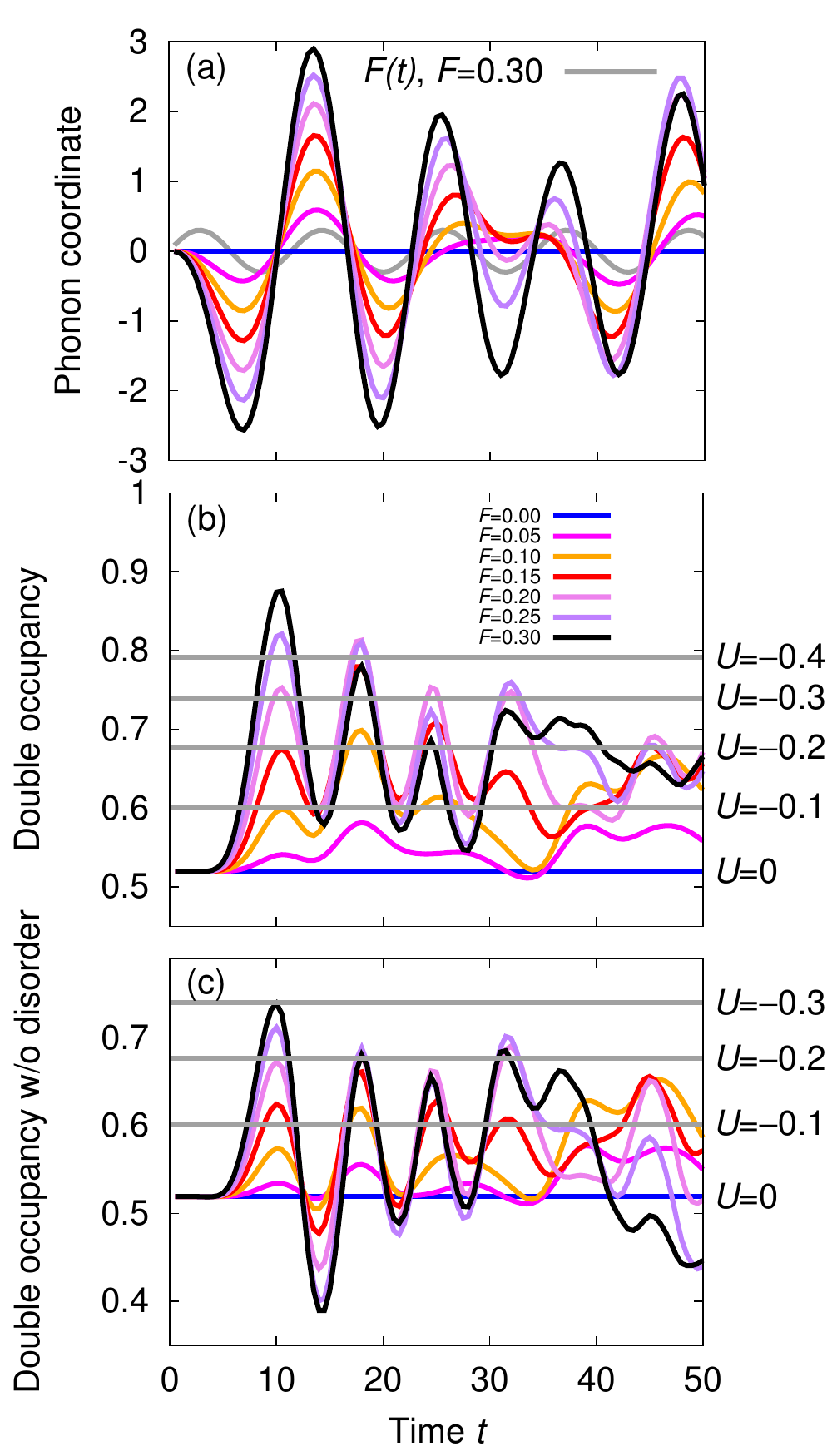}
\caption{
{\bf Light-enhanced double occupancy.} 
(a) Time evolution of the local phonon coordinate $\langle  \hat{x}_1(t) \rangle = \langle  \hat{x}_2(t) \rangle$ for different field strengths, see legend in (b). The grey curve shows the driving field for the case of $F=0.30$. (b) Time evolution of double occupancy $\sum_{l=1,2} \langle \hat{n}_{l,\uparrow}(t) \hat{n}_{l,\downarrow} (t) \rangle$ for different field strengths as indicated. Vertical lines show corresponding equilibrium values for $g_2=-0.05$ and attractive $U$ as indicated. (c) The double occupancy subtracting an effective disorder contribution from the $g_2 \hat{n}_l 2 b^{\dagger}_l b^{}_l$ term, as explained in the main text.}
\label{fig3}
\end{figure}

Having established an increased $\lambda$, we now demonstrate that this also leads to an enhancement of double occupancy, mimicking the effect of light-induced electron-electron attraction. In Fig.~\ref{fig3}(a) the time-dependent driving field is shown together with the time evolution of the local phonon coordinate. Clearly, as discussed above, after a few cycles the phonon shows forced oscillations following the external field, justifying the coherent phonon mean-field picture used above to explain the light-enhanced electron-phonon coupling effect. Then in Fig.~\ref{fig3}(b) we show the double occupancy as a function of time for different field strengths. For noninteracting electrons, the bare value of the double occupancy for the two-site model is 0.5, and it is bounded between 0 and 1. In the undriven case, this value is slightly enhanced since $g_2=-0.05$ already leads to a slight effective attraction. This effective attraction is clearly enhanced in the driven system, and values in excess of 0.8 are achieved for the driving field intensities used here. Some part of this doublon production may be due to the indirect excitation of the electronic subsystem via electron-phonon coupling, but one has to bear in mind that a random electronic configuration has a value of 0.5 on average, which is also the high-temperature limiting value in a thermal ensemble. Therefore double occupancies far in excess of 0.5 indicate that the system is correlated with effective negative $U$ rather than simply excited. 

At this point, however, one should mention that the role of the $g_2$ terms in the Hamiltonian is twofold, as pointed out first in Ref.~\onlinecite{kennes_transient_2017}: (i) there are phonon absorption and emission terms that lead to incoherence and electron-phonon coupling effects; (ii) the local term $g_2 \hat{n}_l 2 b^{\dagger}_l b^{}_l$ acts like an effective ``disorder'' contribution that localizes the electrons. This can be understood by envisioning the phonons not in a number eigenstate of $b^{\dagger}_l b^{}_l$ on all sites, but rather in a coherent state, as is apparently the case (Fig.~\ref{fig3}(a)). Then the electrons ``see'' different onsite potentials on different sites from the phonon number fluctuations, hence a disorder-like potential. The ensuing localization effect also leads to an enhanced double occupancy, as does the effective attraction stemming from the phonon emission and absorption terms. In order to separate the effects, computations were performed without the emission and absorption terms in the Hamiltonian, retaining only the disorder term. The light-induced double occupancy contribution from this disorder contribution was subtracted from the data in Fig.~\ref{fig3}(b), and the resulting net double occupancy from the effective attraction term is shown in Fig.~\ref{fig3}(c). Clearly, there is a net enhancement of the double occupancy that first increases with stronger fields, but then decreases and eventually even drops below the initial value for longer times (black curve). This implies that the light-induced effective attraction effect is eventually suppressed and light-induced localization takes over. This result is consistent with the findings by Kennes et al.~\cite{kennes_transient_2017}.

Overall, the enhancement of double occupancy for moderate field strengths suggests that light-enhanced electron-phonon coupling also leads to effective electron-electron attraction in addition to dynamical spectral weight transfer. For comparison, Fig.~\ref{fig3}(b), (c) also show equilibrium values of the double occupancy computed with $g_2=-0.05$ and an additional onsite interaction $U \sum_{l=1,2} \hat{n}_{l,\uparrow} \hat{n}_{l,\downarrow}$, with negative $U$ values as indicated. A direct connection between enhanced electron-phonon coupling and negative $U$ would require us to integrate out the phonons, which is a standard procedure in equilibrium but requires more care out of equilibrium, especially for driven phonons as is the case here. In fact, a retarded time- and frequency-dependent interaction would naturally emerge, as indicated by the effective self-energy discussed above. Hence more work is required to extract for instance superconducting pairing correlations from out of equilibrium dynamics in the present type of model system.

\subsection{Electronic excitations}
\label{sec:excitations}

\begin{figure}[ht!pb]
\includegraphics[clip=true, trim=0 0 0 0,width=0.9\columnwidth]{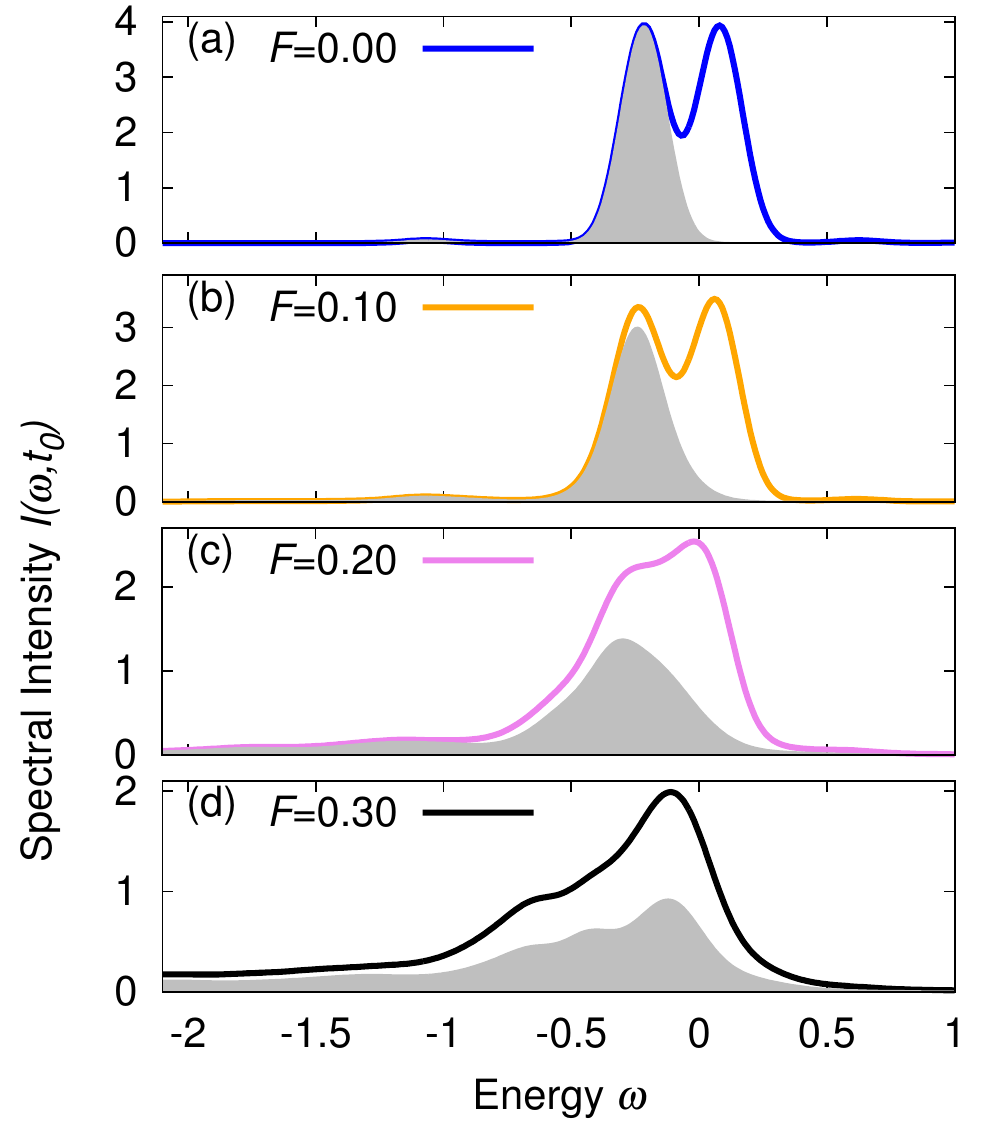}
\caption{
{\bf Electronic excitations.} 
(a) Single-particle spectrum for $g_2=-0.05$ and $F=0.00$. The grey-shaded area shows the occupied part of the spectrum as given by Eq.~(\ref{eq:spectrum}) without the term in the bottom line. (b), (c), (d) The same for $F=0.10$, $F=0.20$, $F=0.30$, respectively.}
\label{fig4}
\end{figure}

Finally, we discuss the role of electronic excitations. This is particularly important when it comes to the possibility of light-enhanced electronic ordering, such as superconductivity, in driven systems \cite{sentef_theory_2016, knap_dynamical_2016, kennes_transient_2017, babadi_theory_2017, murakami_nonequilibrium_2017}. Judging from these references, it is still up for debate whether the enhancement of effective pairing actually leads to light-enhanced or light-induced superconductivity, since for the latter electronic heating effects are important. In order to explore this, we show in Fig.~\ref{fig4} the occupied part of the electronic spectrum (Eq.~(\ref{eq:spectrum}) without the term in the bottom line) that would be measured by time-resolved photoemission spectroscopy. In equilibrium the occupied part follows a Fermi-Dirac distribution at zero temperature (Fig.~\ref{fig4}(a)), as the system is in the groundstate. For weak driving, some excitations are created (Fig.~\ref{fig4}(b)), which become continuously stronger (Fig.~\ref{fig4}(c)). Eventually, for the strongest driving field considered here the antibonding state is strongly occupied (Fig.~\ref{fig4}(d)), indicating a highly excited electronic state. However, we point out that the occupied electronic spectral intensity in the antibonding coherence peak, that would correspond to the conduction band in a solid, grows like a higher power than $F^2$ at weak fields, which should be compared to the $F^2$ scaling of the light-enhanced $\lambda$. Therefore the present model with nonlinear electron-phonon coupling and thus relatively weak electronic excitations is perhaps the most promising candidate to achieve minimal heating together with maximally enhanced attraction. 

\section{Conclusion}
\label{sec:conclusion}

In conclusion, we have shown that light-enhanced electron-phonon coupling can be understood as an effect arising from nonlinear electron-phonon dynamics. We propose that nonlinear electron-phonon coupling should be tested as a possible mechanism to understand transient emergent properties, such as light-induced superconductivity \cite{kaiser_optically_2014, hu_optically_2014, mitrano_possible_2016}, or enhanced electron-phonon coupling seen in THz transport \cite{pomarico_enhanced_2017}. In particular, the predicted linear scaling of the light-enhanced $\lambda$ with the laser field intensity should be checked. Obviously for this induced linear electron-phonon coupling to become relevant, one requires (i) a material with a relatively strong nonlinearity in the electron-phonon interaction for an infrared-active mode, and (ii) high-intensity mid-IR laser pulses to achieve resonant phonon driving with large displacements \cite{forst_nonlinear_2011}. In addition to understanding light-induced superconductivity and guiding efforts to achieve this systematically in new classes of materials, it would also be intriguing to observe other signatures of light-enhanced coupling to the lattice, such as polaronic spectral features in photoemission or optical spectroscopies. Theoretically, searches for materials with strong nonlinearities are in order, as well as studies of the nonequilibrium dynamics in lattice models with such nonlinearities. 

\textit{Acknowledgment.--} Stimulating discussions with A.~Cavalleri, M.~Eckstein, M.~Fechner, I.~Gierz, D.~Kennes, A.~Millis, T.~Oka, E.~Pomarico, and A.~Rubio are gratefully acknowledged. This work was financially supported by the DFG through the Emmy Noether programme.

\bibliographystyle{apsrev4-1}
\bibliography{Nonlinear_coupling}


\end{document}